# Blackout Resilient Optical Core Network

**ZAID H. NASRALLA[1], TAISIR E. H. ELGORASHI[2], and JAAFAR M. H. ELMIRGHANI[2],**

[1]Computer Science Department, College of Science, University of Kerbala, Karbala, 56001, Iraq
[2]School of Electronic and Electrical Engineering, Institute of Communication and Power Networks, University of Leeds, Leeds LS2 9JT, U.K.

Corresponding author: Zaid H. Nasralla (zaid.nasralla@uokerbala.edu.iq)

This work was supported in part by the Engineering and Physical Sciences Research Council (EPSRC), in part by the INTERNET under Grant EP/H040536/1, in part by the STAR Projects under Grant EP/K016873/1 and in part by the TOWS Projects under Grant EP/S016570/1.

**ABSTRACT** A disaster may not necessarily demolish the telecommunications infrastructure, but instead it might affect the national grid and cause blackouts, consequently disrupting the network operation unless there is an alternative power source(s). In this paper, power outages are considered, and the telecommunication network performance is evaluated during a blackout. Two approaches are presented to minimize the impact of power outage and maximize the survival time of the blackout node. A mixed integer linear programming (MILP) model is developed to evaluate the network performance under a single node blackout scenario. The model is used to evaluate the network under the two proposed scenarios. The results show that the proposed approach succeeds in extending the network life time while minimizing the required amount of backup energy.

**INDEX TERMS** —blackout, core network, disaster-resilient, IP/WDM, power outage.

## I. INTRODUCTION

**P**OWER outage is one of the main disruption causes for telecommunication network operations. Although telecommunication networks depend on the power grid, but the authors in [1] described the relation between the Internet and the power grid as an example of infrastructure interdependence as the Internet depends mainly on the power grid to stay operational while the SCADA power control systems use the Internet to communicate.

Natural disasters, man-made disasters or technology faults can cause blackouts. In [2], the researchers classified large scale blackouts as technology-related disasters. During the Japan Earthquake in 2011, the affected area was left in a blackout, where 1500 telecommunications switching offices were left without mains power supply except few limited batteries. Eventually, the switching systems shut down after a few hours when the batteries charge was depleted [3]. In 2005, Hurricane Katrina caused telecommunication disruption due to power outages for 134 networks for ten days [4]. In Italy in 2003 [5], a substation failure led to communication network shut down. Consequently this shut down caused a failure in the SCADA power control system which caused more substations failures and led to a total blackout. This interdependency problem has been studied in a number of papers such as [6]-[8].

Most countries rely on the national grid network, and any disruption in the grid might bring systems in the country as a whole, that rely on electricity, down. The telecommunications Central Offices (COs) are regularly powered by the national grid, while diesel generators and battery cells are used as a backup power supply. Obviously these are of limited availability. Operators are obliged to ensure Business As Usual (BAU) during disasters, though using backup power sources is essential to avoid Service Level Agreement (SLA) violation due to power outages. The recent energy strategies that promote the use of renewable energy can alleviate grid network shut down, however still the use of renewable energy is limited even in developed countries.

Disaster survivability is a trending topic that has been researched extensively within the multi-correlated and large scale failures that happen to nodes and links. In [9], a Software Defined Networking (SDN) approach was developed to mitigate the disaster risk by fast rerouting at the network node and a splicing approach at the controller was used to restore the failed paths. In [10], a disaster resilient virtual network mapping was modelled using a probabilistic approach to evaluate network performance post-disaster to ensure minimal impact on network performance after a single physical link failure. The researchers in [11] proposed mapping virtual networks for SDN controllers, so that any physical infrastructure failure does not compromise the communication between the control and data planes.

In [12], the authors presented Disaster-Resilient Optical Datacenter Networks. They developed an integer linear programming (ILP) optimization model to design an optical datacenter network that considers content placement, routing and protection paths to content. Virtual machines placement across geo-distributed datacenters are studied in [13] to avoid content being isolated in a failed datacenter.

Post-disaster progressive network recovery was investigated in [14], where an ILP was developed to prioritize a recovery plan that considers the restoration of high impact damaged parts. In [15], the availability and the cost of upgrading a damaged core optical network in a post-disaster scenario is considered. The researchers addressed the problem of selecting a set of edges to be upgraded at a minimum cost, while guaranteeing desired values of end-to-end availability through developing geo-diverse routing. In [16], the researchers proposed deploying disaggregated subsystems for rapid post-disaster recovery. Integrating these multi-vendor optical network technologies is possible due to their open interfaces in the control and data planes.

Disasters may have an indirect impact on telecommunication networks, such as huge traffic and power outages. In [17], [18], we have investigated the network traffic floods that are stimulated by disasters. We have studied the network performance with different volumes of traffic, and then suggested four approaches to minimize traffic blocking and serve more traffic using MILP modeling and heuristic approaches. The researchers, in [19], surveyed the impact of recent power outages and the number of affected users. Any of these events would affect not less than a few millions of users. They have suggested the use of Software Defined Networking (SDN) technology to build a Disaster-Resilient SDN Network that can be adaptive to blackout situations. In [20], the authors studied the disasters that cause power outages, and proposed a framework for disasters risk assessment. Based on the assessment, the logistic resources are planned by providing portable generators and permanent solar cells.

Energy-efficiency is another important metric when designing and operating networks under normal conditions and under disaster conditions. It can be considered a mitigation approach for blackouts, as minimizing power consumption can save the limited power sources available. In the energy-efficiency context, we have explored different strategies and approaches to minimize the overall network power consumption [21] – [26]. We considered the use of renewable energy [27], energy efficient physical topology design optimization [28], content placement, caching and replication in the clouds [29] – [31] to reduce power consumption, and energy efficient virtual network embedding [32]. Consideration was also given to the energy efficiency of 5G networks and their functional virtualization [33] as well as developing a power optimized IoT virtualization framework [34]. The use of network coding in IP over WDM networks for energy efficiency was evaluated in [35], [36], while designing cloud and disaggregated datacenters for energy efficiency were presented in [37], [38] and finally green and energy efficient processing of big data was studied in [39] – [42], while [43] studied advance reservation demands scheduling.

Combining these two contexts: energy-efficiency and disaster-resiliency has not been studied before in a power outage scenario. However building an energy-efficient survivable network was investigated in a number of papers, [36], [36], [44] – [46], but the approaches did not consider disaster survivability (multi-correlated failures), large scale disasters or limited power sources. Although power outages were considered in the previous literature but these studies have considered a fully shut down node and how to reroute the traffic after being isolated [2], [35], [44].

In this work, we consider scenarios in which network nodes are powered by the national grid, a renewable power source and backup batteries. The survival time of a blackout node is extended by reducing the amount of traffic that is routed through the limited power nodes. Furthermore, during renewable energy production hours, the available renewable energy is exploited first before using the battery energy. Section II presents the proposed scenarios for building blackout resilient networks. Section III presents the developed MILP formulation for modeling the scenarios. Section IV evaluates the scenarios and discusses the network performance under these scenarios. Section V concludes the paper.

## II. Blackout Resilient Scenarios

We study two scenarios to show the impact of blackouts when building a Blackout-Resilient Network. The first scenario follows the traditional practices that network operators follow in normal and disaster times, while the second scenario is intended for adding resilience to the network to adapt to the blackout situation.

In this work, we consider an IP over WDM core network architecture. Generally, the core network topology is a mesh topology. Therefore, there is more than one path from source to destination.

In the core network, a node failure does not only impact the originated/destined traffic of the failed node but also the transit traffic through the node is disrupted unless there is a protection path for the transit traffic to be rerouted over. Switching the traffic to the protection paths is activated only if the node shuts down. Otherwise if the node is still working, even with limited power, the transit traffic is not rerouted over the protection paths, till the backup energy is depleted.

**Blocking Minimization Scenario**

One of the main KPIs for network performance is the blocking probability. Therefore, network operators are obliged to minimize the expected blocking, that might happen due to traffic anomalies/growth, within the design and operation phases. During the network design, minimizing the expected blocking is typically solved by capacity overprovisioning. The overprovisioning is done by doubling/tripling (or increasing by a larger factor) the network infrastructure resources. These overprovisioned resources are generally put to sleep or standby, and are activated whenever they are needed. In the operation phase, whenever a fault happens in a node or a link, traffic bypassing the faulty node/link should be rerouted on a protection path to avoid traffic disruption. Rerouting the traffic can be done by activating and reconfiguring the overprovisioned resources to serve the rerouted traffic.

The other indicator that network operators look for is the power consumption. Minimizing the power consumption reduces the operational costs, and this is an objective for network operators. Minimizing the consumed power and/or minimizing the blocking probability can be achieved through several approaches during both the design and the operation phases.

One of the intuitive routing approaches that can be employed is the minimum hop routing algorithm [21], [22]. This approach minimizes the operational resources, consequently minimizing the consumed power.
In the blocking minimization scenario, we follow the usual operator practices in network operation by minimizing both blocking and the consumed power, while giving priority to blocking over power consumption.

**Weighted Energy Sources Optimization (WESO) Scenario**
Typically, the Central Office (CO) buildings, that contain the core node, are powered by more than one power source such as grid power, renewable power, and batteries. The power sources have different operational and capital expenditures (OPEX and CAPEX). For example, the grid power has different operational pricing schemes from city to city, or during the day. On the other hand, the renewable power sources have low OPEX, except in terms of preventive maintenance, while such renewable sources typically have higher CAPEX. In contrast, the batteries have limited energy, which means they cannot handle the network operation for long time periods without being charged by another power source. From the operation perspective and according to the above we can conclude that, renewable is preferable to handle network operation as long as it is available, while the grid comes next due to its higher operational costs. Finally, backup batteries comes last, and can be used to power the network for few hours, if both grid and renewable fail.
The above approach can be used by operators to route the traffic in normal times, while in a blackout scenario prioritization should be changed. The change should ensure that the use of grid power is prioritized over the battery and renewable power to preserve the battery energy and the renewable energy to power the blackout node(s). Therefore, this approach attempts to ensure that during a blackout, the available energy is used only to serve the originated and destined traffic of the blackout node, while the transit traffic should be routed away from the blackout node to other nodes where grid power is available. Also the use of renewable energy in the blackout node is given priority over the power drawn from the battery as long as renewable energy is available; to preserve the battery energy to the times when there is no renewable production (i.e. no sunlight or wind).

### III. MILP for Blackout Resilient Scenarios

A MILP model was developed to optimize the routing in core networks under a single node blackout scenario where limited alternative energy sources are available to the blackout node. These energy sources can be renewable sources, batteries or diesel generators used to supplement the national grid power. The objective of the model is to reduce the total power consumption where the batteries power is prioritized in the minimization.

The model considers a bypass IP over WDM architecture which is shown in Figure 1. The available energy sources are assumed to be used to power the network equipment only, i.e. the power consumption of CO cooling system, servers are not considered. In addition, as the focus in this paper is on the core network, the access network and aggregation routers are not considered.

Under the bypass IP over WDM network architecture, the power consumption is composed of:

1- Power consumption of router ports at time t:
$$\sum_{s \in N} \sum_{\substack{j \in N \\ :j=i \lor s=i, \\ s \neq j}} \frac{1}{2} P_r\, C_{sj}$$

2- Power consumption of transponders:
$$\sum_{m \in N} \sum_{\substack{n \in N_m \\ :m=i \lor n=i}} \frac{1}{2} P_t\, W_{mn}$$

3- Power consumption of EDFAs:
$$\sum_{m \in N} \sum_{\substack{n \in N_m \\ :m=i \lor n=i}} \frac{1}{2} P_e\, F_{mn} A_{mn}$$

4- Power consumption of regenerators:
$$\sum_{m \in N} \sum_{\substack{n \in N_m \\ :m=i \lor n=i}} \frac{1}{2} P_g\, RG_{mn} W_{mn}$$

5- Power consumption of optical switches:
$$\sum_{m \in N} P_o$$

Before introducing the model, the parameters and variables used in the model are defined in Table (1).

TABLE I
LIST OF THE SETS, PARAMETERS AND VARIABLES USED IN THE MILP MODEL

| Symbol | Description |
|---|---|
| $N$ | Set of nodes |
| $N_i$ | Set of neighboring nodes of node $i$ |
| $s$ and $d$ | Denote source and destination nodes of a traffic request |
| $i$ and $j$ | Denote end nodes of a virtual link in the IP layer |
| $m$ and $n$ | Denote end nodes of a physical link in the optical layer |
| $TD$ | Time slot duration |
| $P_r$ | Power consumption of a router port |
| $P_t$ | Power consumption of a transponder |
| $P_o$ | Power consumption of an optical switch |
| $P_e$ | Power consumption of an EDFA |
| $P_g$ | Power consumption of a regenerator |
| $B$ | Capacity of a wavelength |
| $W$ | The number of wavelengths per fiber |
| $F_{mn}$ | Number of fibers in link $(m,n)$ |
| $RG_{mn}$ | The number of regenerators in link $(m,n)$ |
| $\lambda^{sd}$ | Traffic request from node $s$ to destination node $d$ |
| $B_i$ | The available battery energy at node $i$ |
| $R_i$ | The maximum output power of the renewable source |
| $\alpha, \beta, \gamma$ and $\delta$ | Weighing coefficients |
| $A_{mn}$ | Is the number of amplifiers between nodes $m$ and $n$, on a link $L_{mn}$, $A_{mn} = \left(\frac{L_{mn}}{s} - 1\right) + 2$, where $s$ is the distance between two neighboring EDFAs and $L_{mn}$ is the distance between nodes $m$ and $n$. |
| $\lambda^{sd}_{ij}$ | The traffic flow of request $(s,d)$ that traverses the virtual link $(i,j)$ |
| $C_{ij}$ | The number of wavelength channels in the virtual link $(i,j)$ |
| $W^{ij}_{mn}$ | The number of wavelength channels in the virtual link $(i,j)$ that traverse link $(m,n)$ |
| $bl_{sd}$ | Binary blocking variable. If $bl_{sd} = 1$ then the request from node $s$ to node $d$ is blocked, otherwise it is not blocked. |
| $RE_s$ | The amount of renewable power consumed at node $s$ |
| $BT_s$ | The amount of power withdrawn from a battery during time slot $t$ at node $s$ |
| $BR_s$ | The amount of grid (brown) power consumed at node $s$ |

The model is defined as follows:

Objective function:

*Minimize*

$$\sum_{i \in N}(\alpha\, RE_i + \beta\, BR_i + \gamma\, BT_i) + \delta \sum_{s \in N} \sum_{d \in N: s \neq d} bl_{sd} \quad (1)$$

The objective function minimizes the power consumed from the different power sources at each node, while keeping the blocking to a minimum. Each power source is weighted by a coefficient. Tuning these coefficients adjusts the operator's energy strategy. The traffic flow conservation can be formulated as:

$$\sum_{j \in N, i \neq j} \lambda^{sd}_{ij} - \sum_{j \in N, i \neq j} \lambda^{sd}_{ji} = \begin{cases} \lambda^{sd}(1-bl_{sd}) & m = s \\ -\lambda^{sd}(1-bl_{sd}) & m = d \\ 0 & otherwise \end{cases}$$
$$\forall s, d, i \in N: s \neq d, \quad (2)$$

Constraint (2) is the flow conservation constraint in the IP layer. It ensures that the total outgoing traffic is equal to the total incoming traffic except for the source and destination nodes.

$$\sum_{s \in N} \sum_{d \in N, s \neq d} \lambda^{sd}_{ij} \leq C_{ij} \quad (3)$$
$$\forall j, i \in N: i \neq j,$$

Constraint (3) is the virtual link capacity constraint. It ensures that the summation of all traffic flows through a lightpath does not exceed the lightpath capacity.

$$\sum_{n \in N_m} W^{ij}_{mn} - \sum_{n \in N_m} W^{ij}_{mn} = \begin{cases} C_{ij} & m = i \\ -C_{ij} & m = j \\ 0 & otherwise \end{cases} \quad (4)$$
$$\forall i, j, m \in N: i \neq j,$$

Constraint (4) is the flow conservation constraint in the optical layer. It assumes that the total outgoing wavelengths in a virtual link should be equal the total incoming wavelengths except the source and the destination nodes of the virtual link.

$$\sum_{i \in N} \sum_{j \in N} W^{ij}_{mn} = W_{mn} \quad (5)$$
$$\forall m \in N, n \in N_m,$$

$$W_{mn} \leq W \cdot F_{mn} \quad (6)$$
$$\forall m \in N, n \in N_m,$$

Constraint (5) finds the total wavelengths per link $(m,n)$, while constraint (6) ensures that the total wavelengths per link do not exceed the fiber link capacity.

$$RE_i \leq R_i \quad (7)$$
$$\forall i \in N$$

$$BT_i TD \leq B_i \quad (8)$$
$$\forall i \in N$$

Constraints (7) and (8) ensure that the power consumed per node does not exceed the available generated energy for renewable and battery sources. Constraint (7) ensures that at each time point the amount of power consumed from renewable sources does not exceed their produced power. In constraint (8) the formulation ensures that the power withdrawn from a battery for the duration of the time slot does not exceed the battery residual energy.

$$\sum_{s \in N} \sum_{\substack{j \in N \\ :j=i \vee s=i, \\ s \neq j}} \frac{1}{2} P_r\, C_{sj} + \sum_{m \in N} \sum_{\substack{n \in N_m \\ :m=i \vee n=i}} \frac{1}{2} P_t\, W_{mn}$$
$$+ \sum_{m \in N} \sum_{\substack{n \in N_m \\ :m=i \vee n=i}} \frac{1}{2} P_g\, RG_{mn} W_{mn}$$
$$+ \sum_{m \in N} \sum_{\substack{n \in N_m \\ :m=i \vee n=i}} \frac{1}{2} P_e\, F_{mn} A_{mn} + P_o \quad (9)$$
$$= BT_i + BR_i + RE_i$$
$$\forall i \in N$$

Constraint (9) makes sure that the power consumed in the node (which consists of power consumed in the router ports, transponders, regenerators, EDFAs and optical switch) should equal the total power withdrawn from all energy sources (batteries, brown and renewable).

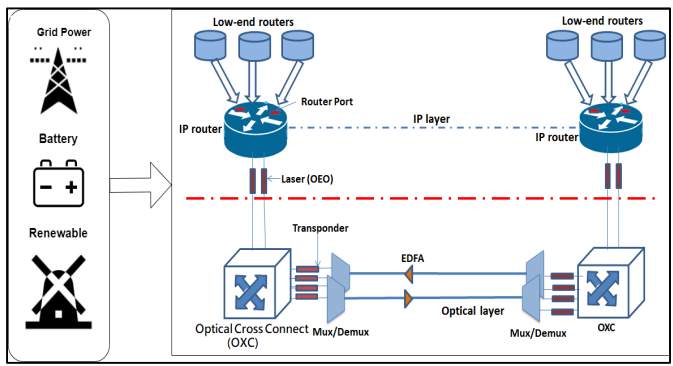

Figure 1 IP over WDM Network Architecture

### IV. Network Performance Evaluation

The model is evaluated using the Italian Network topology shown in Figure 2 which consists of 21 nodes and 36 bidirectional links and one DC located in Milan (node 19). Table II shows the parameters used, in terms of number of wavelengths, wavelength capacity, distance between two neighboring EDFAs, and energy consumption of different components in the network. The average traffic between a node pair varies throughout the day following the profile in Figure 3 with the busiest hour at 22:00. The traffic is generated using a gravity model based on the population of the city where the node is located [21].

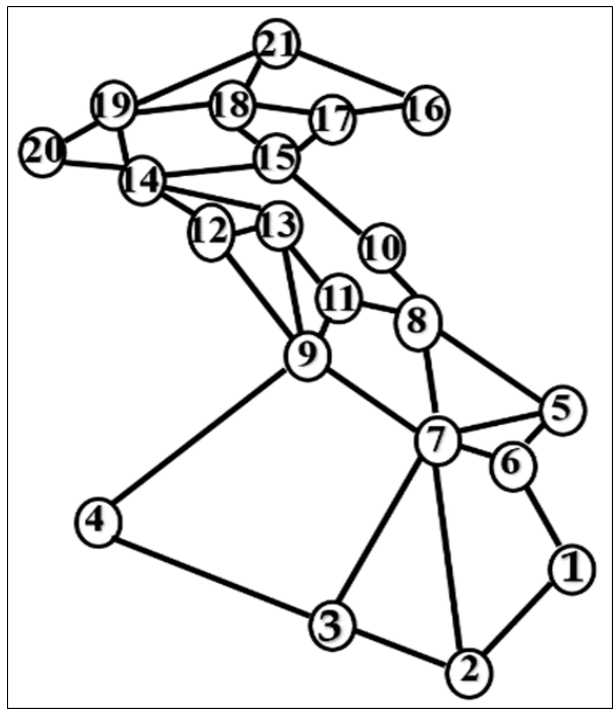

FIGURE 2 Italian Network

TABLE II
NETWORK PARAMETERS

| Parameter | Value |
| --- | --- |
| Distance between two neighboring EDFAs (S) [21] | 80 (km) |
| Number of wavelengths in a fiber (W) [28] | 32 |
| Capacity of a wavelength (B) [22] | 40 (Gb/s) |
| Power consumption of a router port ($P_r$) [21] | 825 (W) |
| Power consumption of a transponder ($P_t$) [21] | 167 (W) |
| Power consumption of a regenerator ($P_g$) [21] | 334 (W) |
| Power consumption of an EDFA ($P_e$) [22] | 55 (W) |
| Power consumption of an optical switch ($P_o$) [29] | 85 (W) |

Solar energy is used as the renewable energy source. Each CO is equipped with 100 m² of solar panels, so the peak solar energy produced in a day is approximately 70 kW as shown in Figure 4, [27], [47]. The sunrise and sunset and the perceived irradiance of April are considered where the sun light is available for 12 hours approximately [48].

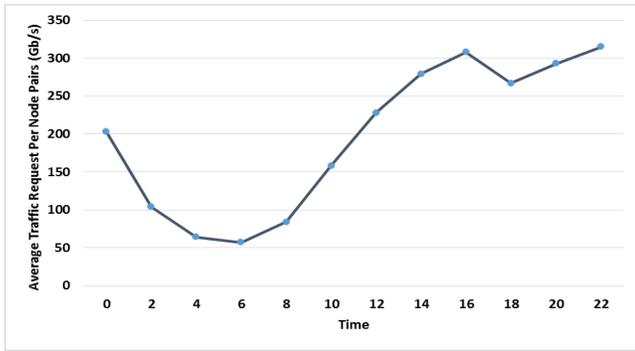

**FIGURE 3.** Average traffic request

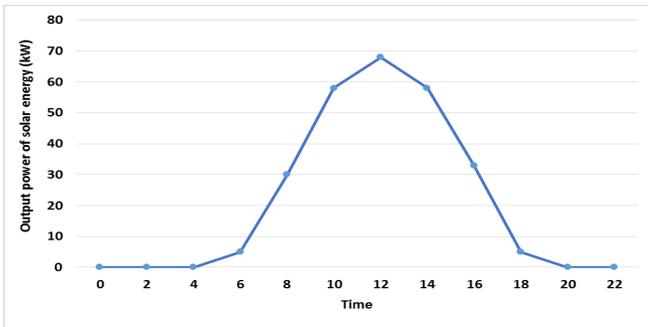

**FIGURE 4.** Solar cells output power at each node

Two optimization approach used the developed MILP model with varying objective function coefficients to study the trade-off between blocking and energy preservation.

*Blocking minimization scenario*: the power consumption coefficients α, β and γ were set to 1 while δ was given a very high number (1,000,000 here), as the main objective is to minimize the total network blocking probability while minimizing the consumed power in total.

*Weighted Energy Sources Optimization* (*WESO*) *scenario*: In this scenario the weights are set in a way that ensures that the use of grid power is prioritized over the battery and renewable power to preserve the battery energy and the renewable energy of the blackout node. Table III shows the three different combinations of weighing coefficients.

The blackout is considered to happen at the beginning of the day (00:00). The model was run for a whole day in a two hour interval in a sequential manner. By assuming that the network receives the requests each two hours and routes them, at the end of the two hours, the residual battery energy is passed over to the next time point and so on up to the end of the day.

TABLE III
WEIGHING COEFFICIENTS

|  | α | β | γ |
|---|---|---|---|
| WESO 1 | 1 | 10 | 100 |
| WESO 2 | 6 | 8 | 20 |
| WESO 3 | 5 | 15 | 25 |

The four proposed scenarios (blocking minimization and three weighted energy optimization) are evaluated for a day long blackout at node 14. The node is assumed to have 360 kWh batteries, (a small saloon, 12 V, car battery is typically rated at 40 ampere hours, which translates to 0.48 kWh. The Tesla electric vehicle battery is 60 kWh to 85 kWh according to car model [49]). Figure 5 shows the power consumed for each power source throughout the day and the battery residual energy for node 14.

In Figure 5(a) the blocking minimization approach results are shown. In this scenario, the node used the battery energy during the midnight hours till 4:00 am, because the traffic is relatively high. Then from 04:00 to 06:00, the node turned off as no energy source is available yet. At 06:00 the sun rises and the node operation is resumed until sunset at 18:00 where the node shuts down again until the end of the day. During the renewable power availability, the node fully utilized the renewable energy generated to serve the node traffic and the transit traffic.

The WESO 1 scenario results are shown in Figure 5(b). The node used minimal energy from the battery energy, because the node handled its own traffic only (originating and destined traffic) while the transit traffic is rerouted. This can be seen as the maximum consumed power in the node ie 14 kW. During the sunlight hours, the node used renewable energy, while after sunset, the batteries were used.

The WESO 2 scenario results are shown in Figure 5(c). The results show that the switching between the batteries and renewable energy is the same as the WESO 1 scenario. During the sunlight hours, however, the node fully used the available renewable energy to route both the destined/originating traffic and transit traffic as the weight given to renewable energy did not stop routing the transit traffic.

The WESO 3 scenario results are shown in Figure 5(d). This scenario behaves similar to WESO 1 and 2 in switching between the batteries and renewable. The only difference is that in WESO 3, the difference in weights between renewable and battery is larger. Therefore less transit traffic is routed through the blackout node resulting in lower blackout node power consumption. This can be verified by checking the total consumed power in the node. In WESO 1, the node consumed 14 kW maximum, while in WESO 2 it consumed 70 kW and in WESO 3 it consumed 28 kW. In conclusion, the three scenarios avoided using the battery energy in two cases. The first when the renewable is sufficient for routing the node traffic. The other situation where the battery power was not used is the situation where the node would have had to route transit traffic under normal conditions when the renewable energy is unavailable. The differences are during the availability of the renewable energy.

The WESO 1 scenario is studied throughout this section, because it avoided the blackout node during the 24 hour blackout. This avoidance comes at the cost of higher power consumption in the network, but to preserve the limited available energy in the node for its own traffic.

To maximize the impact of the blackout, the nodes with the most transit traffic are considered to suffer a blackout. To evaluate the most critical nodes, the MILP model is used assuming that there is no blackout in any node. Figure 6 shows the amount of traffic (transit, originating and destined) carried by each node. Nodes 7, 9, 12, 13, 14 and 19 suffer blackout one node at a time. The two scenarios are evaluated for each node to identify the blocking incurred and how much battery power is required. To compare the blocking, the same battery is considered for the two scenarios. A high capacity battery is assumed which is enough to run the node till the end of the day.

Generally, the nodes can be classified into high traffic nodes and low traffic nodes. The high traffic comes from either a high population in the city, from a DC or from transit traffic passing by the node. According to this classification, nodes 7 and 9 which are in Rome and Napoli have a huge population, while node 19 in Milan has a huge population and there is a DC collocated as well. Nodes 12, 13 and 14 lie in the path leading to the DC. The suggested approaches mainly deal with traffic distribution in the network.

First, nodes 12, 13 and 14 are evaluated for both scenarios considering each node to be equipped with a battery of 360 kWh. This is the battery capacity needed to reduce the blocking probability to zero (i.e. the battery is enough to last the node for 24 hours, for the traffic the node generates and sinks).

Figure 7 shows the blocking probability of these nodes. Clearly, the WESO scenario outperforms the blocking minimization approach as it managed to run the node for the entire day; while under the blocking minimization scenario blocking occurred due to the blackout nodes not being able to send/forward traffic after running out of battery when sunlight is unavailable.

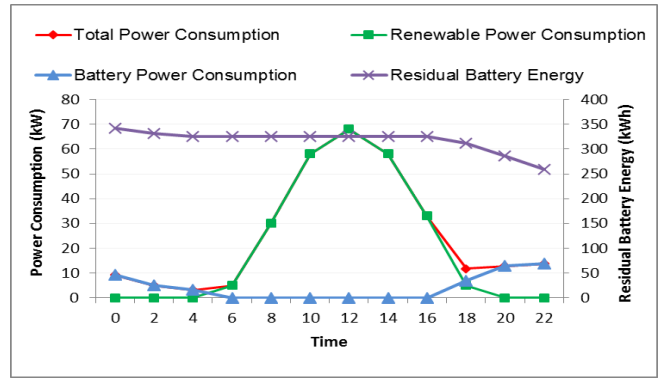

(c)

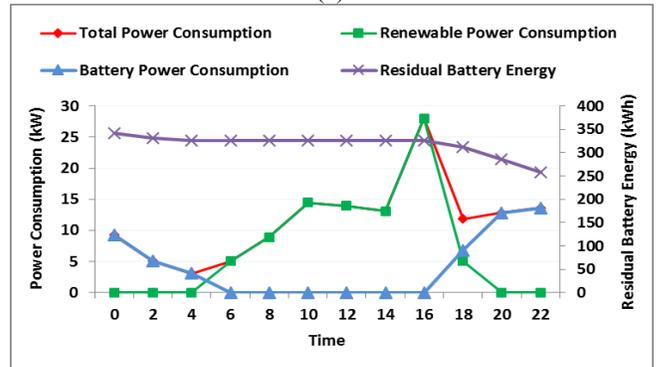

(d)

**FIGURE 5.** Node 14 power consumption and battery residual energy under a) blocking minimization b) WESO1 c) WESO2 d) WESO3

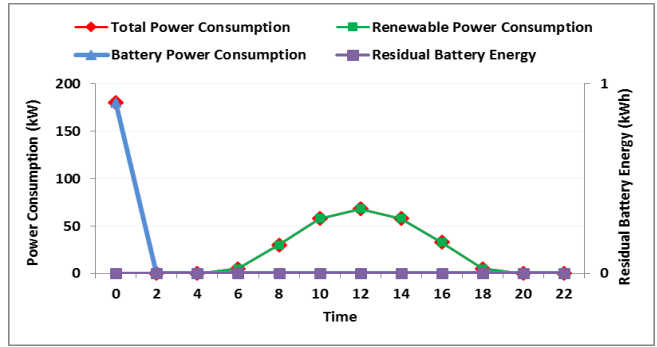

(a)

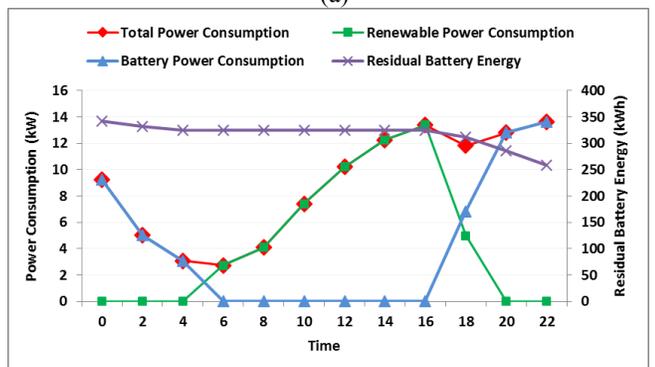

(b)

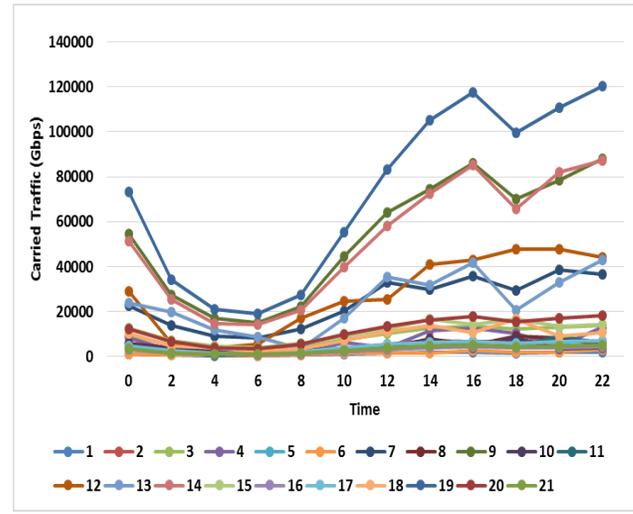

**FIGURE 6.** Node carried traffic through the day

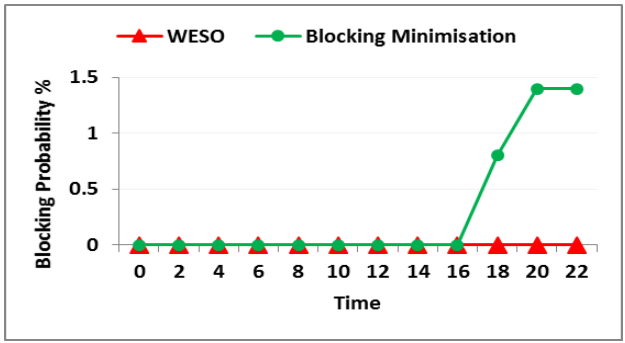

(a) Node 12

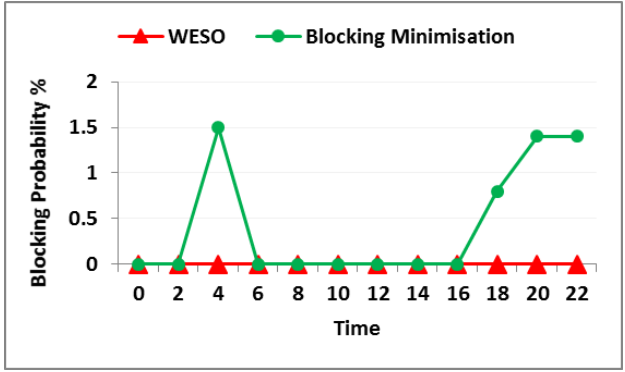

(b) Node 13

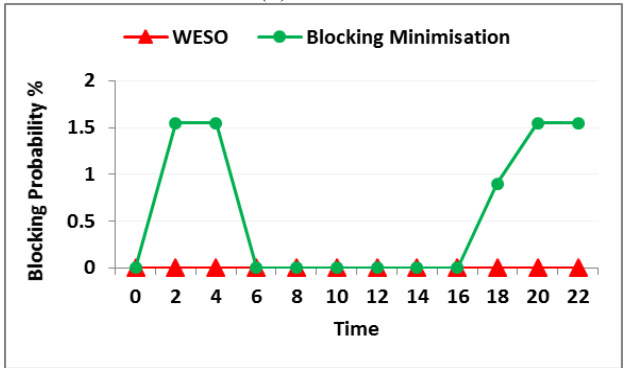

(c) Node 14

**FIGURE 7.** The blocking probability for a) node 12 b) node 13 c) node 14

The blocking minimization scenario used the battery to run the node for the first hours of the day after which the available renewable energy was enough to serve all the demands, then blocking starts when no renewable energy is available. Nodes 13 and 14 start blocking from early hours because all the battery energy is used to carry the transit traffic in the four early hours of the day. The nodes' blocking probability throughout the day as seen in the figure show that node 14 is the worst, then node 13 follows and the least blocking is in node 12. This variation is due to the node running out of battery energy before sunrise. In this sense, node 14 drained the battery energy by 02:00 hour, while node 13 exploited the battery energy from 00:00 to 04:00, node 12 used the battery energy until sunrise. The amount of energy used depends to a large extent, in many cases on the node transit traffic.

Figure 8 shows the total number of hops in three scenarios: WESO scenario, blocking minimization scenario and WESO with no blackout scenario. The no blackout scenario has the lowest number of hops to keep the energy minimized. The WESO scenario rerouted the traffic paths away from the blackout node and this can be seen in the figure as the number of hops stays constant above that of the no blackout scenario. The blocking minimization approach routed based on the minimum-hop paths (because the objective includes minimizing the total power consumption, and minimum hop (not shortest path) minimizes power consumption) during the first two hours of the day till the battery power was exhausted. When the node shut down (02:00 and 04:00), the number of hops decreased due to blocked requests. To minimize the blocking at 06:00, the blackout node was avoided, so the number of hops increased till 20:00 where the renewable power became insufficient to serve the traffic. After sunset, the node has no power to keep it working, so it shuts down, and the number of hops decreased.

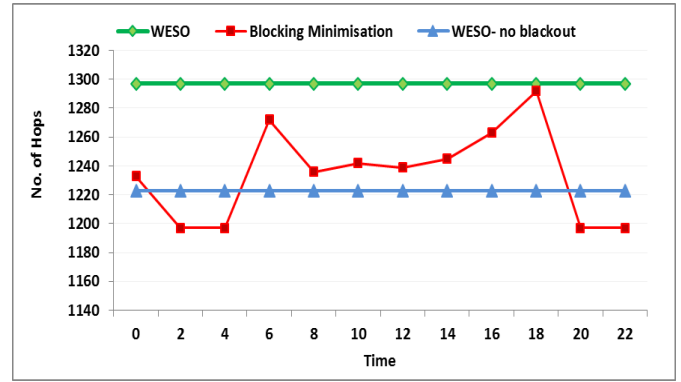

**FIGURE 8.** Number of hops

To determine the battery energy required in node 14 to keep the network running under minimum power consumption, an energy minimization scenario was evaluated using the MILP model without blackout. The results indicated that a 2000 kWh battery energy is required to run the node for 24 hours, while in the WESO scenario the results (in Figure 5) showed that 100 kWh battery energy is sufficient to keep the node working. In conclusion, the blocking minimization approach requires 20 times more battery energy more than WESO scenario. This means more space to store the battery system and more power to keep the battery charged, and it might be infeasible to find the space needed to store this larger system.

Due to their high originating and destined traffic, nodes 7 and 9 are therefore equipped with batteries of 720 kWh and 1500 kWh, respectively. Figure 9 shows the blocking probability for nodes 7 and 9 when a blackout takes place for 24 hours. The WESO scenario has succeeded in keeping the node running for the 24 hours, while the node goes completely out of service during the last four hours of the day in the blocking minimization approach. Under 24 hours blackout at nodes 7 and 9, the blocking minimization approach started blocking earlier than the scenarios with blackouts at nodes

12,13 and 14 because the available renewable energy during the sunlight hours is not enough to serve all the traffic.

Nodes 7 and 9 have huge traffic requests but also they carry the transit traffic between the datacenter (in the north) and the south edge nodes. So the WESO scenario can perform better by rerouting the transit traffic away from them. For datacenter node 19 which lies in the far edge of the network and has originating and destined traffic, the WESO scenario cannot solve the problem.

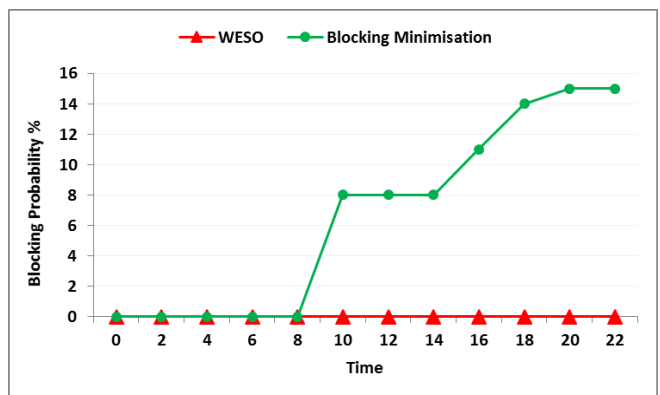

(a) Node 7

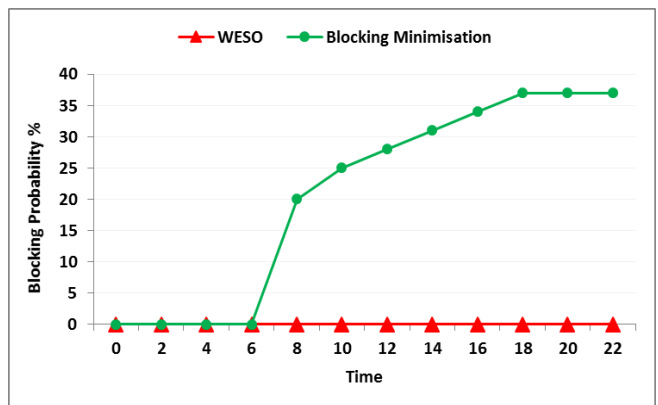

(b) Node 9

**FIGURE 9.** The blocking probability for a) node 7 b) node 9

## V. CONCLUSIONS

Blackouts are a serious source of disruption in the network during disasters unless there is a backup power source. In this paper, building a blackout resilient network has been investigated in the optical core network. Two scenarios have been considered one with the objective of minimizing blocking and the other has the aim of optimizing the usage of power sources where the blackout nodes are considered to have access to solar energy and batteries. A weighted energy optimization scenario was introduced. This attempts to maximize the blackout survival time while minimizing the blocking. A MILP model was developed to optimize the IP over WDM network performance under the two scenarios. An example network was used to evaluate the model with realistic traffic requests. The results show that the WESO scenario succeeded in extending the network life time with the smallest battery resource compared with the blocking minimization approach. Using an online routing technology (or programmable networks) is always a solution such as the SDN. Network topology and node location affects surviving time.

## ACKNOWLEDGMENTS

The authors would like to thank EPSRC for partly funding this work. All data are provided in full in the results section of this paper.

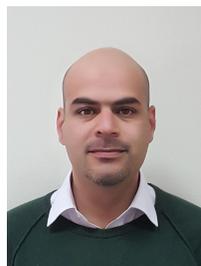

**ZAID H. NASRALLA** received the B.S. degree and the M.Sc. degree in Computer Engineering from Al-Nahrain University, Iraq, in 2002 and 2007, respectively. He received the Ph.D. degree in Disaster Resilient Optical Core Network from the School of Electronic and Electrical Engineering, University of Leeds, U.K., in 2017. He is currently working as a Lecturer and Researcher in the Computer Science Department at the University of Kerbala. His teaching and research interest are in Computer Networks, Network Design, Optical Core Network Technologies, and Building Disaster Resilient and Energy Efficient Networks.

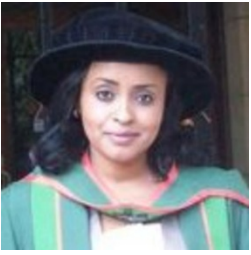

**TAISIR E. H. EL-GORASHI** received the B.S. degree (first-class Hons.) in Electrical and Electronic Engineering from the University of Khartoum, Khartoum, Sudan, in 2004, the M.Sc. degree (with distinction) in Photonic and Communication Systems from the University of Wales, Swansea, UK, in 2005, and the PhD degree in Optical Networking from the University of Leeds, Leeds, UK, in 2010. She is currently a Lecturer in optical networks in the School of Electronic and Electrical Engineering, University of Leeds. Previously, she held a Postdoctoral Research post at the University of Leeds (2010– 2014), where she focused on the energy efficiency of optical networks investigating the use of renewable energy in core networks, green IP over WDM networks with datacenters, energy efficient physical topology design, energy efficiency of content distribution networks, distributed cloud computing, network virtualization and big data. In 2012, she was a BT Research Fellow, where she developed energy efficient hybrid wireless-optical broadband access networks and explored the dynamics of TV viewing behavior and program popularity. The energy efficiency techniques developed during her postdoctoral research contributed 3 out of the 8 carefully chosen core network energy efficiency improvement measures recommended by the GreenTouch consortium for every operator network worldwide. Her work led to several invited talks at GreenTouch, Bell Labs, Optical Network Design and Modelling conference, Optical Fiber Communications conference, International Conference on Computer Communications, EU Future Internet Assembly, IEEE Sustainable ICT Summit and IEEE 5G World Forum and collaboration with Nokia and Huawei.

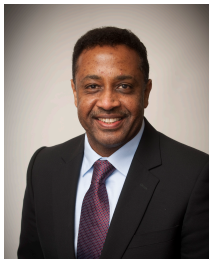

**Jaafar M. H. Elmirghani** is the Director of the Institute of Communication and Power Networks within the School of Electronic and Electrical Engineering, University of Leeds, UK. He joined Leeds in 2007 and prior to that (2000–2007) as chair in optical communications at the University of Wales Swansea he founded, developed and directed the Institute of Advanced Telecommunications and the Technium Digital (TD), a technology incubator/spin-off hub. He has provided outstanding leadership in a number of large research projects at the IAT and TD. He received the Ph.D. in the synchronization of optical systems and optical receiver design from the University of Huddersfield UK in 1994 and the DSc in Communication Systems and Networks from University of Leeds, UK, in 2012. He has co-authored Photonic switching Technology: Systems and Networks, (Wiley) and has published over 550 papers. He has research interests in optical systems and networks. Prof. Elmirghani is Fellow of the IET, Fellow of the Institute of Physics and Senior Member of IEEE. He was Chairman of IEEE Comsoc Transmission Access and Optical Systems technical committee and was Chairman of IEEE Comsoc Signal Processing and Communications Electronics technical committee, and an editor of IEEE Communications Magazine. He was founding Chair of the Advanced Signal Processing for Communication Symposium which started at IEEE GLOBECOM'99 and has continued since at every ICC and GLOBECOM. Prof. Elmirghani was also founding Chair of the first IEEE ICC/GLOBECOM optical symposium at GLOBECOM'00, the Future Photonic Network Technologies, Architectures and Protocols Symposium. He chaired this Symposium, which continues to date under different names. He was the founding chair of the first Green Track at ICC/GLOBECOM at GLOBECOM 2011, and is Chair of the IEEE Sustainable ICT Initiative, a pan IEEE Societies Initiative responsible for Green and Sustainable ICT activities across IEEE, 2012-present. He is and has been on the technical program committee of 41 IEEE ICC/GLOBECOM conferences between 1995 and 2021 including 19 times as Symposium Chair. He received the IEEE Communications Society Hal Sobol award, the IEEE Comsoc Chapter Achievement award for excellence in chapter activities (both in 2005), the University of Wales Swansea Outstanding Research Achievement Award, 2006, the IEEE Communications Society Signal Processing and Communication Electronics outstanding service award, 2009, a best paper award at IEEE ICC'2013, the IEEE Comsoc Transmission Access and Optical Systems outstanding Service award 2015 in recognition of "Leadership and Contributions to the Area of Green Communications", received the GreenTouch 1000x award in 2015 for "pioneering research contributions to the field of energy efficiency in telecommunications", the 2016 IET Optoelectronics Premium Award, shared with 6 GreenTouch innovators the 2016 Edison Award in the "Collective Disruption" Category for their work on the GreenMeter, an international competition, and received the IEEE Comsoc Transmission Access and Optical Systems outstanding Technical Achievement award 2020 in recognition of "Outstanding contributions to the energy efficiency of optical communication systems and networks", clear evidence of his seminal contributions to Green Communications which have a lasting impact on the environment (green) and society. He is currently an editor / associate editor of: IEEE Journal of Lightwave Technology, IEEE Communications Magazine, IET Optoelectronics, Journal of Optical Communications, and is Area Editor for IEEE Journal on Selected Areas in Communications (JSAC) Series on Machine Learning in Communication Networks (Area Editor). He was an editor of IEEE Communications Surveys and Tutorials and IEEE Journal on Selected Areas in Communications series on Green Communications and Networking. He was Co-Chair of the GreenTouch Wired, Core and Access Networks Working Group, an adviser to the Commonwealth Scholarship Commission, member of the Royal Society International Joint Projects Panel and member of the Engineering and Physical Sciences Research Council (EPSRC) College. He was Principal Investigator (PI) of the £6m EPSRC INTelligent Energy awaRe NETworks (INTERNET) Programme Grant, 2010-2016 and is currently PI of the £6.6m EPSRC Terabit Bidirectional Multi-user Optical Wireless System (TOWS) for 6G LiFi Programme Grant, 2019-2024. He has been awarded in excess of £30 million in grants to date from EPSRC, the EU and industry and has held prestigious fellowships funded by the Royal Society and by BT. He was an IEEE Comsoc Distinguished Lecturer 2013-2016.